\documentclass{nature}
\usepackage[usenames,dvipsnames]{color}
\usepackage{graphicx}
\usepackage{dcolumn}
\usepackage{bm}
\usepackage{textcomp}

\def\figtxt{}
\def\mcla{}
\def\mcl{}

\def\nc{n_{\mathrm{c}}}

\def\ls{\ell_{\mathrm{s}}}
\def\lplasma{\ell_{\mathrm{p}}}
\def\ni{n_{\mathrm{i}}}
\def\ne{n_{\mathrm{0}}}
\def\nh{n_{\mathrm{e}}}
\def\refl{\xi_{\mathrm{2}}}
\def\fah{f_{\mathrm{e}}}
\def\fap{f_{\mathrm{i}}}
\def\gamh{\gamma_{\mathrm{e}}}

\def\bh{\beta_{\mathrm{e}}}
\def\bhvec{\mathbf{\beta_e}}
\def\bpz{\xi_{\mathrm{1}}}

\def\falim{f_{*}} 
\def\rh{\xi_{\mathrm{3}}}

\def\lc{\ell_{\mathrm{c}}}
\def\fmax{f^*}

\def\il{I_\mathrm{l}}
\def\laml{\lambda_\mathrm{l}}
\def\kl{\mathbf{\hat{k_l}}}

\title{Petawatt laser absorption bounded}

\author{Matthew C. Levy$^{1,2}$, Scott C. Wilks$^2$, Max Tabak$^2$, Stephen B. Libby$^2$ \& Matthew G. Baring$^1$}

\begin{document}

\maketitle

\begin{affiliations}
 \item Department of Physics and Astronomy, Rice University, Houston, Texas 77005, USA
 \item Lawrence Livermore National Laboratory, Livermore, California 94551, USA
\end{affiliations}

\begin{abstract}
\mcla{The interaction of petawatt ($10^{15}\ \mathrm{W}$) lasers with solid matter forms the basis for advanced scientific applications such as table-top particle accelerators, ultrafast imaging systems and laser fusion. Key metrics for these applications relate to absorption, yet conditions in this regime are so nonlinear that it is often impossible to know the fraction of absorbed light $f$, and even the range of $f$ is unknown. Here using a relativistic Rankine-Hugoniot-like analysis, we show for the first time that $f$ exhibits a theoretical maximum and minimum. These bounds constrain nonlinear absorption mechanisms across the petawatt regime, forbidding high absorption values at low laser power and low absorption values at high laser power. For applications needing to circumvent the absorption bounds, these results will accelerate a shift from solid targets, towards structured and multilayer targets, and lead the development of new materials.}
\end{abstract}

\section{Introduction}

Irradiation of solids by petawatt laser light ($\il\ \laml^2 > 10^{18}\ \mathrm{W\ \mu m^2 \ cm^{-2}}$, where $\il$ is intensity and $\laml$ is wavelength) creates extreme states of matter with temperatures exceeding ten million degrees Celsius and pressures exceeding one billion earth atmospheres. 
These high energy density conditions are driven at the microscopic scale by dense currents of relativistic electrons ($\sim 10^{11} \   \mathrm{A\ cm^{-2}}$), oscillating violently in the intense laser fields ($> 10^{10}\ \mathrm{V\ cm^{-1}}$), as well as the plasma processes arising when these particles are dephased and injected into the high density target.\cite{Macchi2013a}  
\mcla{
Suitably harnessed, this setup opens the way to 
table-top relativistic particle accelerators,\cite{Macchi2013a,Pukhov2003a,Wilks1997,Daido2012,Robinson2013,Hegelich2006,Fuchs2005a,Toncian2006,Esirkepov2004,Bulanov2010c,Roth2002,Sentoku2003,Flippo2008a,Petrov2009,Kar2012} 
laser fusion,\cite{Tabak1994,Kodama2004,Theobald2006,Johzaki2011,Bartal2011,Robinson2014,Norreys2014,Kemp2014} 
laboratory astrophysics,\cite{Ryutov1999,Remington2006,Kugland2012a} 
ultrafast imaging systems,\cite{Wilks2001,Barty2004,Zylstra2012} 
high-energy radiation sources\cite{Nakamura2012} 
and intense high harmonic generation.\cite{Norreys1996,GibbonBook,Gordienko2005,Dollar2013} 
Over the past two decades, the promise of these applications has driven considerable theoretical and experimental study of the crucial problem of how the laser energy is converted to target particle energy.
}
\mcla{Dozens of energy transfer mechanisms have been identified,\cite{Wilks1997,Macchi2013a,Pukhov2003a} and most treatments to date have focused on examining individual mechanisms in isolation to help guide interpretation of results.
In realistic situations, however, these absorption mechanisms can be strongly nonlinear and several often act concurrently.}

\mcla{In this article we report the theoretical maximum and minimum absorption for each laser-solid configuration across the petawatt regime. We find that these extrema constrain nonlinear absorption mechanisms,\cite{Wilks1997,Brunel1987,Kruer1984,Naumova2009,Kemp2008,May2011}
  bounding the laser energy transfer in a more general manner.}
  The present analysis overcomes difficulties of particle nonlinearity by creating a kinematic basis on which to formulate the interaction.
We use a geometry centered at the laser-matter interface, taking advantage of the laser decay into an evanescent wave over a relativistic collisionless skin depth in the optically-thick target. 
Here Rankine-Hugoniot-like conservation laws\cite{Taub1948,DeHoffmann1950} must be satisfied by the forward-going evanescent light wave, \mcla{the} backward-going reflected wave, and forward-drifting highly relativistic electrons and moderately relativistic ions accelerated by the laser.
By representing the complex motion of individual particles with ensemble properties such as density and momentum, accounting for the relativistically-correct laser-solid physics,\cite{Wilks1992,Haines2009,Ping2008,Robinson2009,Gibbon2012,Kemp2012,Levy2013PoP} we realize an essentially four body kinematics situation.
We show that these kinematics \mcla{restrict values} the ensemble properties of electrons and ions can take on.
Since acceleration of electrons and acceleration of ions are modes of absorption of laser light, we demonstrate that these kinematic restrictions can be transformed into useful upper and lower bounds on absorption.
Excellent agreement with a broad range of published experimental and simulation data\cite{Naumova2009,Ping2008,Levy2013PoP,Davies2009} confirms that the absorption bounds are distilling a fundamental aspect of the nonlinear dynamical physics.
\mcla{
For applications using solid targets, our results show a new general metric for measuring efficiency. 
Since the design space to be explored is contracted, these findings will enable research efforts to focus on useful regions of parameter space thus accelerating the development of future laser-solid applications.
We also identify applications requiring efficiency exceeding that permitted by the absorption bounds. 
Our results indicate that these applications would benefit by shifting towards structured\cite{Zhao2010,Jiang2014,Purvis2013} or multilayer\cite{Sgattoni2012,Sarri2013} target designs.
}

\section{Results}

\subsection{\mcla{Relativistic interaction model}}
Essential features of petawatt laser-solid interactions are shown in \figtxt{Fig. 1.}  
Here an ultraintense $a_\mathrm{0}>1$ light pulse (where $a_\mathrm{0} = e E_\mathrm{l}/(m_\mathrm{e}\ c\ \omega_\mathrm{l})$ is the  laser strength parameter, $c$ is the speed of light, $e$ is the fundamental charge, $m_\mathrm{e}$ is the electron mass, $E_\mathrm{l}$ is the laser electric field and $\omega_\mathrm{l}$ is the laser angular frequency)  
is seen to irradiate a thick target of electron density $\ne(x) > \nc$, for realistic spatial profile $\ne(x)$ and critical density $\nc = m_\mathrm{e} \omega_\mathrm{l}^2/(4 \pi e^2)$ (see Methods for additional details). Electrons oscillate relativistically in the intense laser fields allowing the light wave to penetrate into the field-ionized overdense \mcla{(optically-thick)} plasma\cite{Cattani2000} an axial distance equal to the Lorentz-transformed collisionless skin depth, $\ls = \gamh^{1/2}\  \times\ c/\omega_{\mathrm{pe}}$ (where $\gamh=(1-\bh^2)^{-1/2}$ is the electron Lorentz factor, $\bh c$ is the electron speed and $\omega_{\mathrm{pe}}= (4 \pi e^2 \ne / m_\mathrm{e})^{1/2}$ is the plasma frequency).
This forms the scale size for the interaction. Within $\ls$, radiation reaction effects are small due to Debye shielding and electron and ion collisional mean-free paths satisfy $\lambda_{\mathrm{mfp}} \gg \ls$. 
Therefore, as shown in \figtxt{Fig. 1}, electrons and ions are the only particle populations entering into the petawatt-scale kinematic interaction.
Irrespective of the specific mechanism of energy transfer, these particles absorb energy from the laser collisionlessly,\cite{Wilks1992,Denavit1992} and their ensemble properties enter into formulae describing the total absorption $f = \fap + \fah$, where $\fap$ is the absorption into ions and $\fah$ is the absorption into electrons. 

Unbounded $\fah$ and $\fap$ solutions are obtained by applying a relativistic kinematic model at the laser-matter interface, establishing a connection between the laser pulse and the particles it excites across the density discontinuity\cite{Levy2013PoP} (realizing an essential similarity to the Rankine-Hugoniot relations in magnetohydrodynamic shocks\cite{DeHoffmann1950}).
Ion dynamics are constrained by a snow plow-like process called 'hole punching' driven by the laser ponderomotive pressure, which can exceed  $> 10^9$ atmospheres.\cite{Wilks1992}  Electron dynamics, on the other hand, can be governed by a number of different collisionless mechanisms depending on parameters such as laser polarization and angle of incidence.\cite{Wilks1997,Brunel1987,Kruer1984,Naumova2009,Kemp2008,May2011}
In order to calculate results independent of the specific mechanism, ensemble electron properties are determined based on a general Lorentz-invariant ansatz distribution function.\cite{Kluge2011}
Solutions accounting for these realistic dynamical conditions are computed numerically, however an analytic form exists for the representative case $\bhvec \cdot \kl \approx 1$ for laser propagation in $\kl$ (see Methods). Here ion absorption is $\fap = 2 \bpz \refl^{3/2} \  / \ [ \ \sqrt{\bpz^2 \refl+1}-\bpz \sqrt{\refl} \ ]$ and electron absorption is 
$\fah = [ \ (1-\refl) \sqrt{\bpz^2 \refl+1} -  (1+\refl) \bpz \sqrt{\refl} \ ] \ / \ [ \ \sqrt{\bpz^2 \refl+1}-\bpz \sqrt{\refl} \ ] + O\left( \ \rh \ / \ \bpz^2 \ \right)$, using the convenient control parameters $\bpz, \refl$ and $\rh$. Intensity and density conditions are controlled by $\bpz = \left[ \ Z m_\mathrm{e} \nc \ /  \ (2 M_\mathrm{i} \ne) \ \right]^{1/2} a_\mathrm{0}$ for uniform interface charge state $Z$ and ion mass $M_\mathrm{i}$.  $\refl  = I_{\mathrm{l,reflected}}/ I_{\mathrm{l,incident}}$ corresponds to net photon flux deposited in the laser-matter interface, and $\rh = \nh m_\mathrm{e}/[\ni(M_\mathrm{i}+Z m_\mathrm{e})] \ll 1$ is a small parameter exhibiting the disparate mass scales that characterize the petawatt laser-solid absorption modes.

\subsection{\mcla{Absorption bounds}}
\mcla{Absorption bounds reflect the fact that} solutions to the kinematic equations for $\fah$ and $\fap$ can become nonphysical for values of $f$ between zero and one.  These bounds are derived using constrained optimization techniques\cite{Mehrotra1992} with $f=\fap+\fah$ as the objective function.
We optimize $f$  over $\refl$ imposing the simple constraint that the electron energy is real, and the minimization equation is written as $\falim  =  \mathrm{Min} \left( \fah + \fap \right)$, $\ s.t.\ \ \gamh \ge 1$.
Because the utility function is nonlinear in the control variables, minimization is performed numerically by means of  cylindrical algebraic decomposition\cite{Collins1975}, and the resulting points are fit to a polynomial using interval $\Delta \bpz \approx 10^{-3}$ over the physically-relevant range in $\bpz$ between $[0.01, 0.5]$.
For the fully-ionized laser-plasma interface we calculate that $\falim \approx 1.9 \bpz - 2.75 \bpz^2 + 1.91 \bpz^3$, indicating that the lower limit on laser absorption is closely related to the process of ion acceleration by an intense circularly-polarized radiation pressure source\cite{Wilks1997,Naumova2009,Schlegel2009}. Deviations from the absorption associated with this process occur at small $\bpz$  as energy is reapportioned into relativistic electrons, highlighting that the kinematic coupling between ions and electrons represents an important feature of the interaction.  When the electronic coupling is removed, we confirmed that $\falim$ converges to the well-established ion acceleration result $\falim \rightarrow 2 \bpz/(1+2\bpz)$.\cite{Naumova2009}  Maximizing the absorption through $\fmax  =  \mathrm{Max} \left( \fah + \fap \right)$, again subject to the constraint that $\gamh \ge 1$, computes the upper limit to be $\fmax=1-\rh/(2\ \bpz^{2})$. In contrast to the lower limit, there is no well-established analytic result that describes absorption along the $\fmax$ curve based on a simple physical mechanism.
Here we proceed allowing that $\nh \approx \nc$, implying that absorption along $f=\fmax$ corresponds to electrons excited with $\gamh \approx a_0^2 / 2$, within a factor of order unity of the full laser ponderomotive potential.
\figtxt{Fig. 2} presents a comprehensive description of the absorption, showing surfaces corresponding to $\fah$ and $\fap$, as well as bounding regions corresponding to $\falim$ and $\fmax$. 
Ions are seen to dominate the absorption along $f=\falim$, the region corresponding to $\gamh=1$. As the target absorbs more of the laser energy,  \figtxt{Fig. 2} shows that this energy is predominantly coupled into relativistic electrons.   Electrons dominate the absorption along $f=\fmax$, with $f>\fmax$ causing $\fah$ to take on complex value. 

\subsection{\mcla{Comparison between absorption bounds and published data}}
Comparison to published data is facilitated by specifying an interaction-averaged density, which is well-represented by a corrected relativistic critical density\cite{Cattani2000} given by $a_0^2 \approx (27/64) \left(\ne/\nc\right)^4$  for $\ne/\nc \gg 1$.
Fig. \figtxt{3} shows these limits applied to experimental data and simulation results published over two decades, spanning a broad range of laser and plasma parameters, obtained at several laser facilities.   
\mcla{From this figure it is clear that} experimental data and kinetic particle-in-cell simulations at a variety of realistic conditions\cite{Naumova2009,Ping2008,Levy2013PoP,Davies2009} show excellent agreement with the absorption bound predictions.

\subsection{\mcla{Absorption bounds in terms of laser and plasma parameters}}
Transforming from the control coordinates, the condition $\falim \leq f \leq \fmax$ can be written simply in terms of the laser \mcla{power} and unperturbed plasma density as,
\begin{eqnarray} 
      \frac{ \sqrt{ \il \laml^2 } }{\sqrt{ \il \laml^2 } + 1.5 \sqrt{ \ne }}
   \ \leq \ f \ \leq \ 
1 - \frac{1.2 \times 10^{18}}{\il \ \laml^2}
\label{eqn:faIneq}
\end{eqnarray}
where $[ \il \ \laml^2 ] = \mathrm{W\ \mu m^2 \ cm^{-2}}, [ \ne ] =  \mathrm{cm^{-3}} $, and $\il \lambda^2_l > 1.3 \times \ 10^{18}$. 
Equation (\ref{eqn:faIneq}) bounds the laser-solid interaction through its dynamical history for a realistic time-dependent laser envelope and plasma profile.\cite{Naumova2009,Levy2013PoP} 

\section{Discussion}

Fig. \figtxt{2} and \figtxt{3} highlight that \mcla{$\falim$} becomes increasingly strict with laser power, forbidding 35\% of possible absorption values at $\il \ \laml^2 \sim 10^{22}\ \mathrm{W\ \mu m^2 \ cm^{-2}}$, a regime accessible at laser facilities such as ELI\cite{ELI} scheduled to come online in the next few years. \mcl{These results will therefore play a central role in guiding the next generation of multi-petawatt experiments.}


For applications needing to circumvent the absorption bounds in equation (\ref{eqn:faIneq}), these results will drive a shift towards new interaction paradigms.
In order to see that assumptions underpinning the laser-solid interaction model must be violated in order to exceed these limits it is instructive to examine the two outlying points shown in \figtxt{Fig. 3}.
The data point labeled $^{(\dagger)}$ corresponds to one simulation of a pre-deformed, very thin target of $\laml > d$ where $d$ is the target thickness, 
realizing a strongly refluxing configuration (see Methods).  The point labeled  $^{(\dagger\dagger)}$ corresponds to one simulation of a laser interacting with $20 \mathrm{\mu m}$ of  $\nc > \ne$ plasma in front of a thin $\ne=20 \nc$ target, realizing an essentially underdense situation.
We thus confirm that very thin and underdense targets allow absorption in excess of $\fmax$ at low laser power, as they should.  
However, \mcla{several important applications that have conventionally used solid targets} also depend on high absorption at relatively low laser power.
These applications include laser-based anti-matter generation for scaled astrophysical studies,\cite{Liang1998,Chen2009d} ultrafast charged-particle imaging systems,\cite{Wilks2001} where increasing absorption reduces noise and improves imaging resolution, and certain approaches to electron-driven fast ignition laser fusion.\cite{Tabak1994}
Recently works that have shifted from \mcla{solid targets} have started to report enhanced results in these areas.\cite{Purvis2013,Sgattoni2012,Sarri2013} The results presented here will accelerate this shift across the petawatt field,
and lead the development of novel low density, structured and multilayer targets.

\begin{methods}
The essential kinematic relations forming the basis of the optimization analysis reported in this \mcla{article} were published in Ref. \cite{Levy2013PoP}.  \mcl{The general consideration of optimal couplings under the constraint of phase space conservation motivates the present studies.\cite{Fisch1993}}
Radiation-hydrodynamic simulations show that \mcla{particle density in interactions can often be approximated} by an exponential distribution $\ne(x) \propto e^{-x/\lplasma}$ for scalelength $\lplasma$, due to amplified spontaneous emission (ASE) associated with laser pulse compression generating a 'pre-plasma.'\cite{Macchi2013a} 
Petawatt laser-solids satisfy $\lplasma \ [\mathrm{\mu m}] < 1.1 \ a_0 \ \tau_\mathrm{l}\ [\mathrm{ps}]$ for pulse duration $\tau_\mathrm{l}$ such that the primary interaction occurs in the classically-overdense $\ne/\nc > 1$ region while small-scale underdense regions are swept away by the strong laser ponderomotive force, as indicated by energy balance between electron acceleration in the underdense and overdense regions.
The laser temporal envelope  $\il\ (\partial \il/\partial t)^{-1} \gg 2\pi \ \omega_{\mathrm{pe}}^{-1}$ and the plasma density profile is subject to $\ne\ (\partial \ne/\partial x)^{-1} \gg \ls$, both readily satisfied under realistic conditions.
Damping of transient momentum effects requires that $\tau_\mathrm{l}\ \omega_{\mathrm{pi}} > 2 \pi A$ where \mcla{$\omega_{\mathrm{pi}}= (4 \pi e^2 Z \ne / M_\mathrm{i})^{1/2}$} and $A\simeq 3-5$, and the target thickness \mcl{$d$} should exceed the hole punching depth and the effective refluxing hot electron range, $d > c\ \tau_\mathrm{l}/2 + \int_0^{\tau_\mathrm{l}} V_{\mathrm{int}}\ dt$ for motion of the laser-matter interface at velocity $V_{\mathrm{int}}$.
Deviations from $\bhvec \cdot \kl \approx 1$ are second order in the angle $\tan^{-1} \ [\ |\bhvec \times \kl|/ \bhvec \cdot \kl \ ]$ and therefore do not substantively affect the absorption bound results for realistic scenarios.  Energy apportionment into ions increases with this angle but qualitative trends in $\fah$ and $\fap$ are maintained.\cite{Levy2014Ifsa}
For electrons following the ponderomotive energy scaling\cite{Wilks1992}, the cooling length $\lc$ associated with the radiation reaction force within $\ls$ can be estimated\cite{Esarey1993} as $\lc \approx 2.1 \times 10^{-2} \sqrt{ \ne [\mathrm{cm^{-3}]} }\ a_0^{-7/2} \ \ls$ due to Debye shielding. Thus, $\lc/\ls \sim 1$ only at the 10-petawatt level $\sim 10^{23}\ \mathrm{W\ \mu m^2\ cm^{-2}}$, confirming the selection of absorption modes characterizing the petawatt-scale interaction.
\end{methods}


\begin{addendum}
  \item[Acknowledgements] M. L. is grateful to B. Breizman and A. Link for discussions.
M. L. acknowledges the LLNL Lawrence Scholarship for support and the Institutional Computing Grand Challenge for computational resources. 
\mcla{S. W. acknowledges support from Fusion Energy Sciences in the DOE Office of Science.}
This work was performed under the auspices of the U.S. Department of Energy by Lawrence Livermore National Laboratory under Contract DE-AC52-07NA27344. 
\item[Author Contributions] 
M. L. developed the analytical theory with support from S. W. and M. T.
Assistance with conceptualizing the physical arguments and preparing the manuscript was provided by S. L.
Theoretical support was provided by M. B. 
\item[Competing Financial Interests] 
\mcla{The authors declare no competing financial interests.}
 \item[Author Information] 
 Reprints and permissions information is available at www.nature.com/reprints.
 
 Correspondence and requests for materials
should be addressed to M. L.~(email: levy11@llnl.gov).
\end{addendum}

\clearpage

\begin{figure}
\begin{center}
\includegraphics[width=0.9\linewidth]{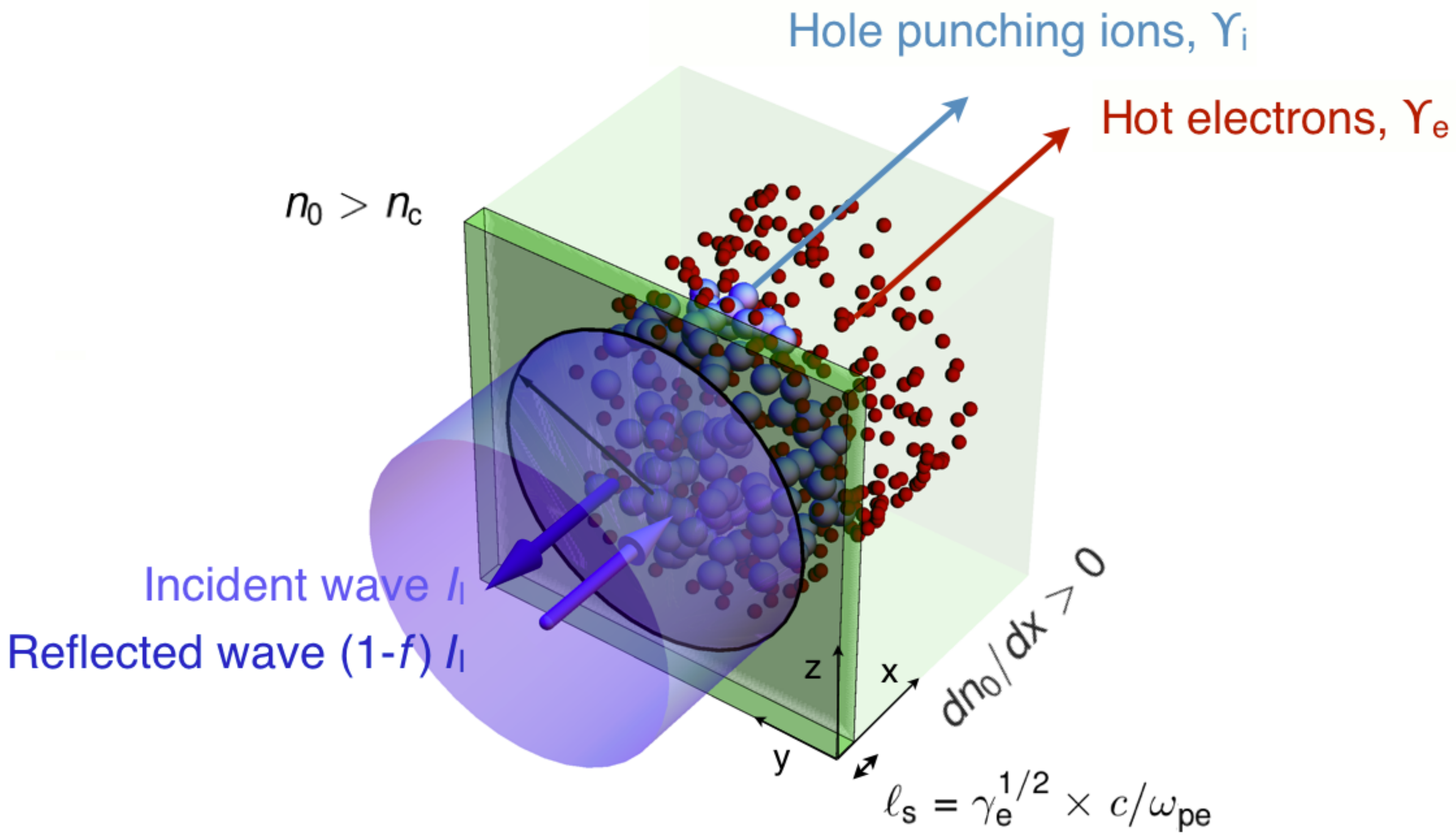}
  \end{center}
  \caption{
  \textbf{Schematic showing key features of the petawatt laser-solid interaction} 
  A high-power laser with strength parameter $a_0>1$ is shown striking an overdense target, interacting over the Lorentz-transformed collisionless skin depth $\ls$ (dark green region), and exciting a highly-relativistic electron flux (red spheres) and moderately-relativistic ion flux (blue spheres).  
  Laser and excited particle properties are connected across $\ls$ by applying relativistic Rankine-Hugoniot-like relations at the laser-matter interface, allowing abstraction of downstream effects, e.g., scattering in the $x>\ls$ target (light green region).
  Depiction uses a frame of reference co-moving with the interface.
  }
  \label{fig:schem} 
\end{figure}

\clearpage

\begin{figure}
\includegraphics[width=0.9\linewidth]{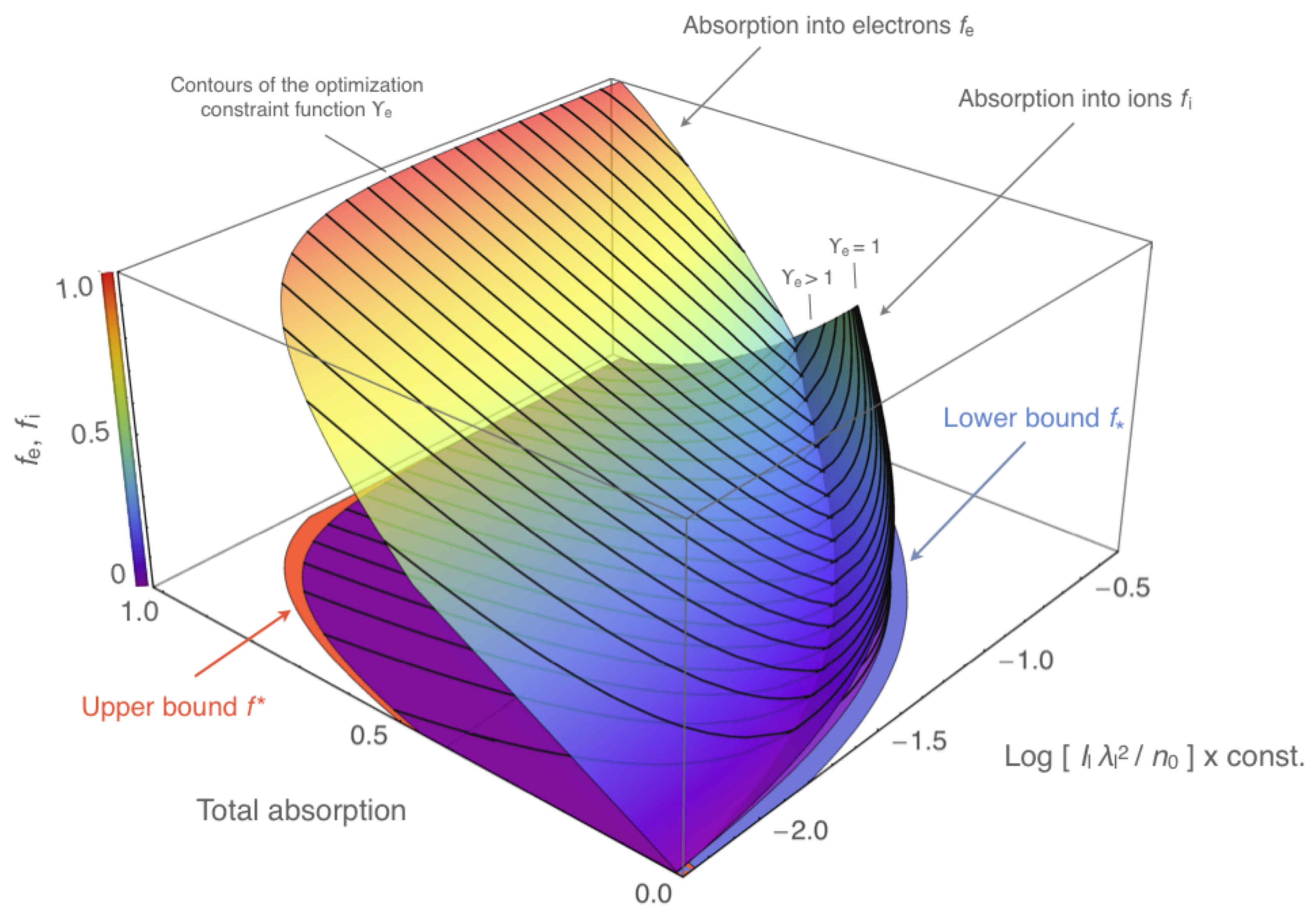}
\caption{
\textbf{Bounds on absorption and sub-partitioning of absorbed light} 
The lower-right axis corresponds to $\bpz$ describing the initial laser and target conditions, and the total absorption $f=1-\refl$.
Two surfaces corresponding to the absorption into electrons $\fah$ and into ions $\fap$ are shown (the former having slight transparency for visualization purposes).  
Contours of the optimization function $\gamh$ are superimposed on these surfaces using dark gray. The lower limit $\falim$ (blue) and upper limit $\fmax$ (red) on absorption are shown bounding $\fah$ and $\fap$.
}\label{fig:phase}
\end{figure}

\clearpage

\begin{figure}
\begin{center}
\includegraphics[width=0.9\linewidth]{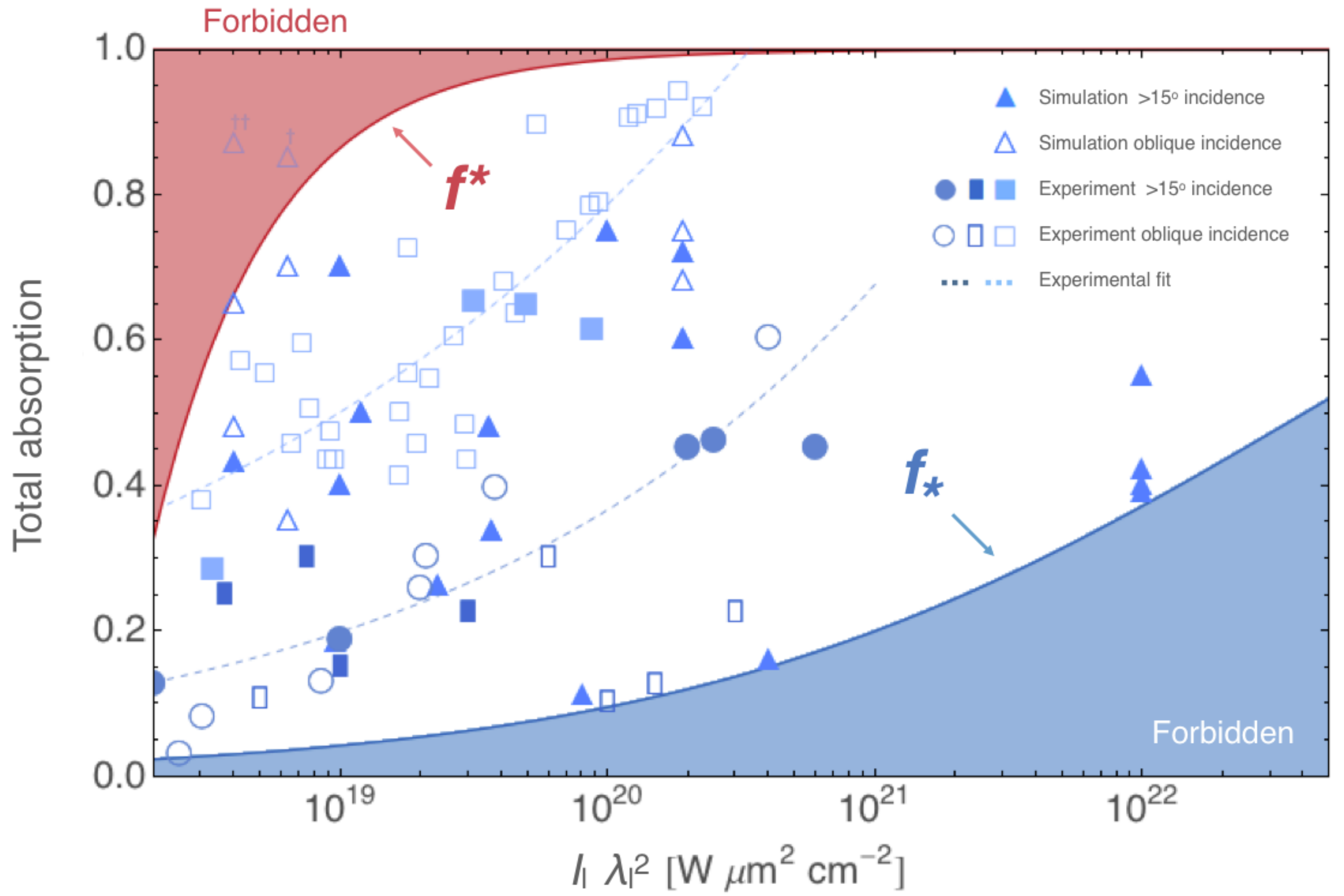}
  \end{center}
  \caption{
  \textbf{Comparison between \mcla{absorption bounds} and published data}
      The complete dataset compiled in \textit{Davies}\cite{Davies2009} is reproduced here, spanning experimental and simulation data published over the past two decades, across a variety of laser and plasma conditions.   Dashed lines corresponding to fits of selected experimental data are shown to guide the eye.  Additional high-intensity simulation data is reproduced from \textit{Levy et al.}\cite{Levy2013PoP}
   The upper limit on absorption $\fmax$ is depicted in red and the lower limit $\falim$  in blue, with forbidden regions indicated using shading. 
   The two outlying data points correspond to simulations of $^{(\dagger)}$ a very thin $0.2 \mathrm{\mu m}$ pre-deformed target, and $^{(\dagger\dagger)}$ an essentially underdense $\nc > \ne$ interaction, both violating assumptions underpinning the laser-solid model (see Methods). 
  }
  \label{fig:limit} 
\end{figure}


\end{document}